\documentclass[useAMS,usenatbib]{mnras}
\usepackage{float,graphics,epsfig,array,amsmath,amssymb}
\usepackage{times,deluxetable,lipsum, hyperref}

\def\aj{AJ}%
\def\araa{ARA\&A}%
\def\apj{ApJ}%
\def\apjl{ApJ}%
\def\apjs{ApJS}%
\def\ao{Appl.~Opt.}%
\def\aap{A\&A}%
\def\aapr{A\&A~Rev.}%
\def\mnras{MNRAS}%
\def\ssr{Space~Sci.~Rev.}%
\def\nat{Nature}%
\def\gca{Geochim.~Cosmochim.~Acta}%
\def\refbf{}

\title
[Light Echos: NGC 4051]
{Measuring Light Echos in NGC 4051}

\author
[T.~J.~Turner, L.~Miller, J.N.Reeves,V.Braito]
{T.~J.~Turner$^{1}$,
L.~Miller$^{2}$,
J.N.Reeves$^{3,4}$,
V.Braito$^{3,5}$ \\
$^{1}$Dept. of Physics, University of Maryland Baltimore County, Baltimore, MD 21250, U.S.A. \\
$^{2}$Dept. of Physics, Oxford University, Denys Wilkinson Building, Keble Road, Oxford OX1 3RH, U.K. \\
$^3$ Center for Space Science and Technology, University of Maryland Baltimore County, 1000 Hilltop Circle, Baltimore, MD 21250, USA \\
$^{4}$Astrophysics Group, School of Physical and Geographical Sciences, Keele 
University, Keele, Staffordshire ST5 5BG, U.K \\
$^{5}$ INAF-Osservatorio Astronomico di Brera, Via Bianchi 46, I-23807 Merate (LC), Italy\\
}

\date{Accepted 2016. Received 2016; in original form 2016}

\pubyear{2016}

\begin{document}
\label{firstpage}
\pagerange{\pageref{firstpage}--\pageref{lastpage}}
\maketitle

\begin{abstract}

{\refbf
Five archived X-ray observations of NGC~4051, taken using the {\it
  NuSTAR} observatory, have been analysed, revealing} 
lags between flux variations in bands {\refbf covering a wide range} of X-ray
photon energy. In all pairs of bands compared, the harder band
consistently lags the softer band by at least 1000\,s, at temporal
frequencies $\sim 5 \times 10^{-5}$\,Hz.  In addition, soft-band lags
up to 400\,s are measured at frequencies $\sim 2 \times
10^{-4}$\,Hz.  Light echos from 
{\refbf an excess of soft band emission in} 
the inner accretion disk cannot explain the lags in these data,
as they are seen in cross-correlations {\refbf with energy bands} 
where the softer band is expected to have no contribution from reflection.  
{\refbf The basic properties of the
  time delays have been parameterised by fitting a top hat response
  function that varies with photon energy, taking fully into account the covariance between measured
  time lag values.  The low-frequency hard-band lags and the transition to soft-band lags
  are consistent with time lags arising as reverberation delays from
  circumnuclear scattering of X-rays, although greater model complexity is required to explain the
  entire spectrum of lags. } The scattered fraction
increases with increasing photon energy {\refbf as expected}, and 
the scattered fraction is
high, indicating the reprocessor to have a global covering fraction
$\sim 50\%$ around the continuum source.  
{\refbf
Circumnuclear material, possibly associated with a disk wind at a few hundred gravitational radii
from the primary X-ray source, may
provide suitable reprocessing}.

\end{abstract}


\section{Introduction}

Variations in X-ray flux are common in Active Galactic Nuclei (AGN),
where factors of several change are evident over all timescales
observed, i.e.  from tens of seconds to years.  
Interpretation of the
flux variations and the dependences on photon energy may allow insight
into the physical processes that are important in the 
immediate neighborhood of the
supermassive nuclear black hole.
Construction of a power spectral density (PSD) can help characterize
the variations, showing the relative importance of the different
temporal modes sampled by the data, while the lag spectrum shows the time lags between
two light curves extracted from different energy ranges;  
these are analyzed as a function of the frequency of Fourier
modes (convention defines a positive lag as
hard photons lagging soft). 
  
Time delays are expected to arise from reverberation in the X-ray band,
caused by reprocessing and scattering of the continuum radiation by circumnuclear material. 
Of particular interest is the so-called transfer function, that is a measure of
the time delays between time series
constructed in differing bands of X-ray photon energy, analyzed as
a function of the temporal frequency of modes of variation. Energy- and 
frequency-dependent time delays
appear to be a common feature of the X-ray emission from AGN
\citep[e.g.][]{demarco13a}.

There are currently two principal mechanisms that have been invoked in
the literature to explain time delays between X-ray energy bands. The
first of these is time delays that may be associated with the possible
inwards propagation of fluctuations on the accretion disk
\citep[e.g.][]{kotov01a, arevalo06a}. The inner accretion disk is
thought to be hotter than the outer disk, so a delayed fluctuation
arriving at the inner disk may stimulate harder X-ray emission than
would be produced at larger radii, leading to an energy-dependent time
delay of hard photon flux variations lagging soft photon flux
variations.  However, while this model may work well for X-ray
binaries with their hotter accretion disks, in AGN the observed X-ray
emission is not thought to arise directly from the disk, as the disk
should be too cool. Instead, it is commonly supposed that disk photons
are Compton upscattered in energy in a hot corona, of unknown geometry
\citep{haardt91a}. 
However, in order to produce the observed powerlaw
X-ray spectrum, the optical depth to Compton scattering must be
sufficiently high that memory of the input spectrum is largely lost
\citep{titarchuk94} - thus it seems likely that this model needs to assert
that coronal fluctuations may produce the require spectral-dependent
time delays.
It is also commonly supposed that the X-ray source is a compact
region close to the black hole event horizon, in order to create a high
equivalent width of strongly
redshifted Fe\,K line emission \citep{miniutti04a}. However, 
if the coronal size is too small then there is no clear mechanism for the 
inwards disk fluctuations to be transferred into a change in the 
X-ray emission via the corona: the 
corona has to be extended in order for the
inwards propagation model to work.
Despite
these difficulties, the `propagation of fluctuations' model remains a popular choice in the
literature \citep{uttley14a}.

The second possible mechanism supposes that time delays arise from light travel
time delays, arising when X-ray photons Compton scatter from circumnuclear
material.  Such a mechanism has been invoked to explain either short timescale
delays in AGN \citep[e.g.][]{fabian09a} or the entire transfer function of
delays across all timescales \citep[e.g.][]{miller10a}. These possibilities will
be discussed later in this paper in the context of new observations of the nearby
AGN NGC\,4051.

In the context of reverberation models, the delayed signal, or ``light
echo'' arises from scattering of the X-ray continuum
from absorbing circumnuclear or accretion disk material - a process
commonly known as `reflection' in the case of optically-thick material.
The reverberation component is expected to have a hard
spectrum relative to the primary continuum and the contribution of
reflection increases with energy. An observed time series generally
comprises both direct emission and delayed, scattered emission, so that
the net observed reverberation time delay is diluted according to the fraction of direct
light that is present. 
In the harder energy bands the contribution of delayed emission increases
and observed time lag tends towards an undiluted value \citep{miller10a}.
Reverberation models predict we should see a strong delayed signal in
the hard X-ray band, peaking above 10\,keV.

Owing to low sensitivity, lack of spatial resolution and high
background levels, there has been no mission prior to {\it NuSTAR}
that has allowed useful reverberation analysis above 10\,keV.  Early results
show the value of {\it NuSTAR} in X-ray reverberation analysis.
\citet{zoghbi14a} presented the first analysis of high-frequency time
lags in {\it NuSTAR} data above 10\,keV, showing the Compton hump to
lag the continuum variations by 1--2\,ks in MCG--5-23-16.
\citet{kara15a} claimed a similar result for SWIFT\,J2127.4+5654 and
NGC\,1365. Emission from around Fe\,K$\alpha$ was also found to lag the
continuum in all three sources.

This work presents a study of the timing and spectroscopic properties
of NGC 4051, observed in five {\it NuSTAR} observations. In 
Section\,\ref{background}
we discuss recent analyses of X-ray observations of NGC 4051. 
In Section\,\ref{data}, we describe the {\it NuSTAR} data reduction. 
In Section\,\ref{spectral} we present the spectral analysis of the
source.
In Section\,\ref{timedelays} we describe our timing analysis methods
and in Section\,\ref{discussion} we discuss the results in the context of
a light echo model for the X-ray reprocessor.

\section{NGC 4051}\label{background}

NGC\,4051 is a bright, nearby narrow-line Seyfert\,1 AGN with low
black hole mass, (M$\simeq 1.7 \times 10^6 {\rm M_\odot}$; \citealt{Denney10a}) 
and high accretion rate ($\sim 10$\,\% of Eddington), 
making it one of the most variable Seyfert galaxies in the X-ray band,
with factors up to 15 observed on timescales of tens of ks
\citep{ponti06a}.  The X-ray spectrum has been well studied and shows
a multi-phase outflow, with zone velocities ranging from a few hundred
to few thousand km/s, \citep{pounds04c, pounds11a}. The ionization of some of
the absorbing zones has been observed to change in response to the
varying continuum \citep{lobban11a, krongold07a}. The strong and
persistent reprocessing signatures show that there is a large amount
of circumnuclear material in NGC 4051, making this an ideal target for
a detailed reverberation study.

Outflowing reprocessor zones with velocities $50-400$\,km\,s$^{-1}$
are also detected in UV spectroscopy of this source \citep{kaspi04a}.
These fit well into the \citet{krongold07a} picture of an absorbing
multi-phase wind and it may be that a significant part of the observed
reflection comes from such a wind \citep{turner07a}.  The X-ray
continuum itself varies by up to a factor 15 on timescales as short as
20\,ks \citep{ponti06a}.  In its low state O{\sc vii} and O{\sc viii}
emission lines from distant photoionized gas are prominent
\citep{pounds04a,ponti06a} and there is likely also a component of
distant neutral reflection \citep{ponti06a}.  Ionised Fe {\sc x}
emission is also detected optically up to 150\,pc from the central
source \citep{nagao00a}. In summary, the reprocessed spectrum for
NGC~4051 appears to include contributions from the inner accretion
disk, the broad-line region and from more distant dusty regions of
circumnuclear gas.

To study the temporal behaviour of the X-ray emission, \citet{mchardy04a} combined 6.5 years of
observations by {\it RXTE} with a long observation by {\it XMM-Newton} (hereafter {\it XMM}), finding the
lag of hard X-ray fluctuations to increase and coherence to
decrease as band separation increases. They also found the lag
and coherence to be greater for variations of longer time
period. This behaviour suggested higher photon energies and shorter
variability timescales to be associated with smaller radii.

\citet{miller10a} presented two long {\it Suzaku} exposures of
NGC\,4051, finding a strong hard component above 10\,keV, some
fraction of which varies with the power-law continuum. The power
spectrum was non-stationary, showing differences between the low and
high flux states. This was confirmed in {\it XMM} observations 
by \citet{alston13a}, who also
measured negative time lags and 
found that the shape of the lag spectra depends on source flux.
Frequency-dependent positive time lags 
$970 \pm 225$\,s were measured by \citet{miller10a} and suggested to be
explained by the effects of reverberation in the hard band,  caused by reflection from a thick shell of material with maximum lags of about 10,000s. 
The reflecting material must 
extend to a distance of about $1.5 \times 10^{14} {\rm cm}$, {\it
  i.e.} 600 gravitational radii from the illuminating source  with high  
global covering factor  ($C_g > 0.44$).  The reflection may occur from
Compton thick parts of the disk wind, seen out of the line-of-sight.
These very
long lags in the hard band also exclude the possibility that the hard component
originates as reflection from the inner accretion disk in NGC~4051 \citet{miller10a}.

\section{Analysis of the {\it NuSTAR} data}\label{data}

\begin{table*}
\centering
\caption{Observation Log}
\begin{tabular}{| l | c | c | c | c | c | c |}
\hline

Date & 
Observation & 
Exposure & 
ct/s &
ct/s &
flux  &
flux \\
 & 
 & 
 & 
2-10\,keV &
10-50\,keV &
2-10\,keV  &
10-50\,keV \\
\hline
2013-06-17 16:41:07 & 60001050002 &  9.4 & 0.25 & 0.11 & 1.99 & 3.79 \\
2013-06-17 21:21:07 & 60001050003 & 45.7 & 0.25 & 0.11 & 1.80 & 3.17 \\
2013-10-09 13:41:07 & 60001050005 & 10.2 & 0.19 & 0.09 & 1.31 & 2.89 \\
2013-10-09 20:01:07 & 60001050006 & 49.6 & 0.13 & 0.07 & 0.93 & 2.29 \\
2014-02-16 13:36:07 & 60001050008 & 56.7 & 0.35 & 0.13 & 2.99 & 4.19 \\
\hline
\end{tabular} \\
\vspace{-0.7cm}
\tablecomments{Observed fluxes are given for FPMA in units 10$^{-11}$ erg\,cm$^{-2}{\rm s^{-1}}$}. 
\label{tab:data}
\end{table*}%

{\it NuSTAR} observed NGC 4051 during five observations spread over
2015 May 15 -- 22, with exposure times between 9.4 -- 56.7 ks
(Table\,\ref{tab:data}).  The data from both focal plane modules (FPMA and FPMB) and
for all observations were reprocessed and cleaned using the most
recent version of the {\it NuSTAR} pipeline (within HEASOFTv16.6) and
calibration files (CALDB version 20140715), resulting in 172 ks of
good on-source exposure (Table\,\ref{tab:data}).

We extracted spectra and light curves from a circular region of 30''
radius, centered on the nucleus.  Background spectra and light curves
were extracted from two circular regions that are free of any obvious
point sources and the ratio of source to background areas for the
extraction cells was 1:5.5.

NGC\,4051 shows strong X-ray variability across the observation set
(Table\,\ref{tab:data} and Fig.\, 1).  The background is 0.5-2\% of the total counts in
the 2-70\,keV band, 2-5\% in the 10-70\,keV band.  There is no concern
regarding dead time in this count rate regime (Table\,\ref{tab:data}).

Examination of the background-subtracted light curves reveals strong
time variability within all of the observations (Fig\, 1). As
observations 60001050002 and 60001050003 are very close in count rate
and spectral shape and were taken close together in time, we combine
those exposures into a single spectrum for the spectral analysis
(Table\,\ref{tab:params}).

The spectral data from those summed observations, plus spectra from
the other three observations were fit for modules FPMA and FPMB
simultaneously.  The calibration offset recommended between those two
instruments is in the range 0.95-1.05, and so a constant component was
allowed in the model, linked to be a single floating value that we fit
to obtain a cross-normalization constant 1.05 between the two FPM
instruments.

\section{Spectral fitting}\label{spectral}

\begin{figure}
\begin{center}
\hbox{
\psfig{figure=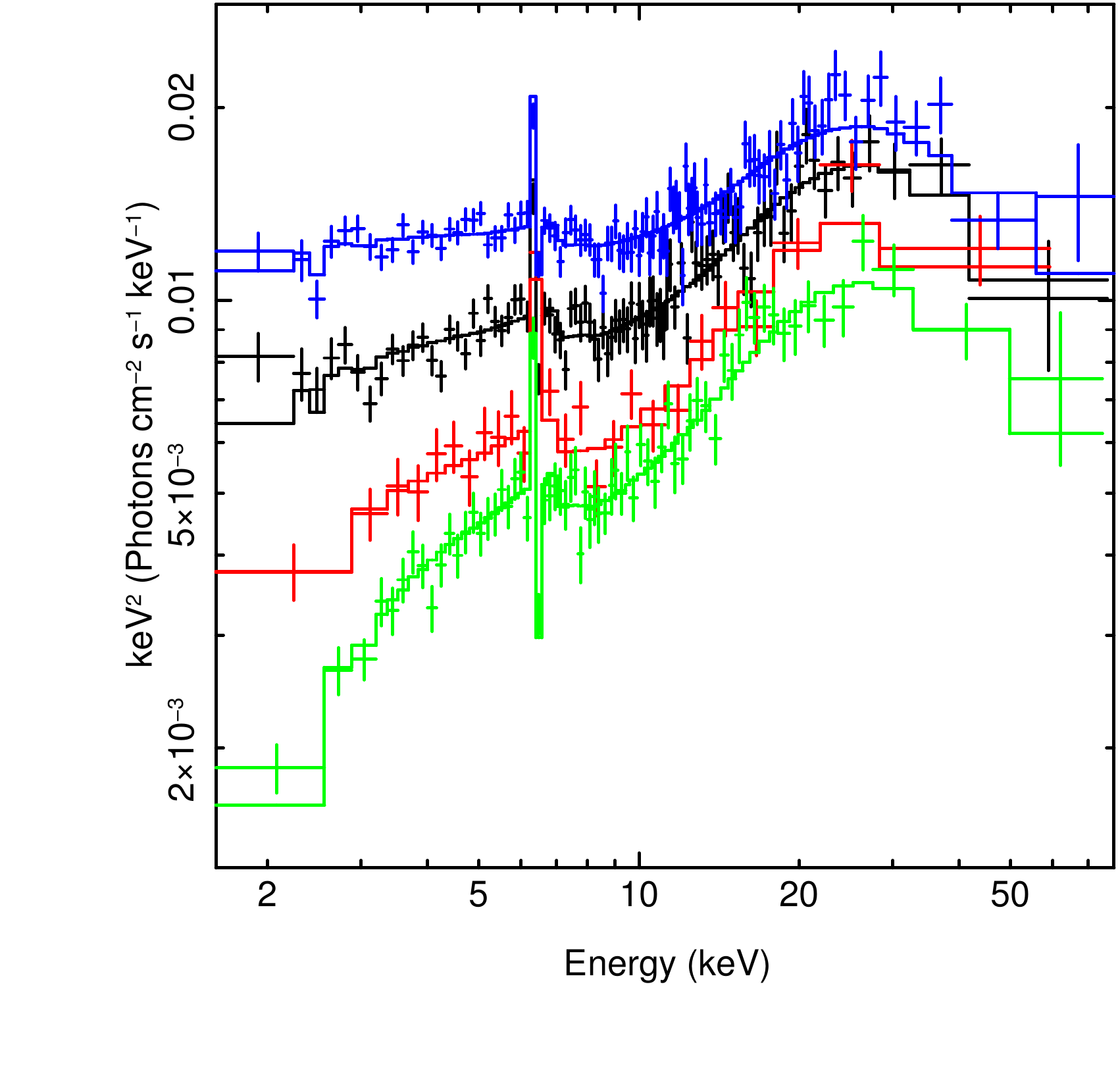,height=1.9in,width=2.4in, angle=0}
}
\end{center}
\vspace{-0.7cm}
\caption{  
{\it NuSTAR} data for NGC\,4051 - for clarity, only FPMA data are shown. Spectral variability is dominated by changes in the continuum level that drive lagged changes in scattered light.  Observations correspond to the colored lines as: 02$+03$ (black), 05 (red), 06 (green) and 08 (blue) -- see Table\,\ref{tab:params}}
\label{fitspectra}
\end{figure}

\begin{figure}
\resizebox{0.5\textwidth}{!}{
\includegraphics{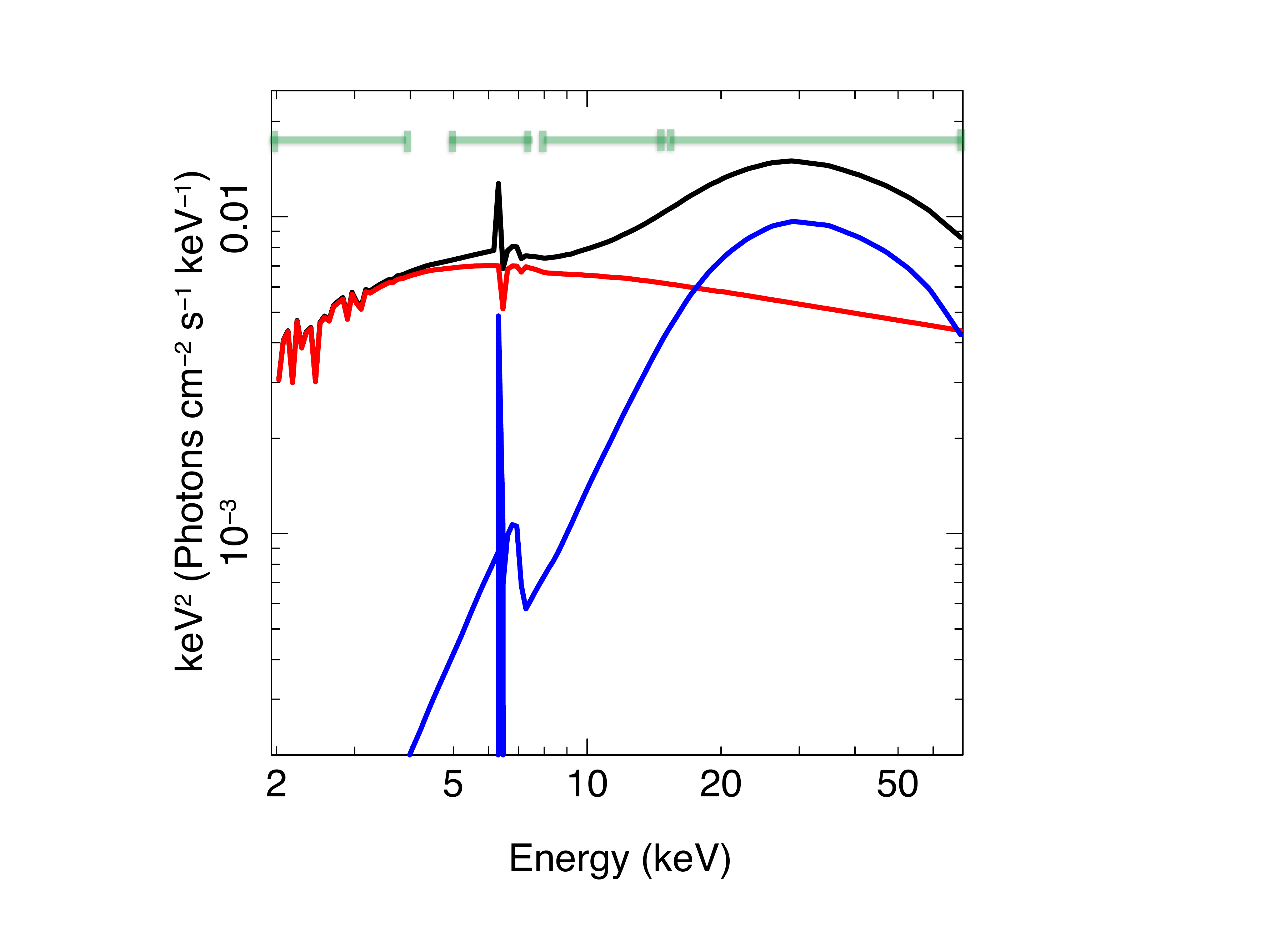}
}
\caption{The mean model for NGC 4051 with the boundaries shown for the bands chosen for timing analysis shown as green horizontal lines. The mean powerlaw component is shown in red, the mean scattered component is shown in blue. Both components are modified by a layer of ionized gas and the summed model is shown as a black line.}
\label{modelplot_bands}
\end{figure}

\begin{table*}
\centering
\caption{Reprocessing Model} 
\begin{tabular}{| l | l | l | l | l | l |}
\hline
Observation &
${\rm N_H}^1$ &
N$_{\rm PL}^{2}$ & 
N$_{\rm Sc}^{3}$ & 
F$_{\rm Fe}^4$  &
EW$_{\rm Fe}$ \\
\hline
02$+$03  &  4.75$^{+1.25}_{-1.19}$    & $12.31 \pm0.9$          &  $7.06\pm0.15$      & 2.78$\pm 0.74$  & 144 \\
05           &  6.28$^{+2.12}_{-2.50}$     & $7.84\pm0.66$            & $6.72\pm0.15$         &    $4.82\pm0.11$ & 347 \\ 
06           & 12.75$^{+1.95}_{-4.11}$    & 6.91$\pm0.52$           & $5.05\pm0.10$     &  1.67$\pm0.65$ & 191 \\
08           & 2.70$^{+0.80}_{-0.70}$       & 17.33 $\pm1.17$       & $6.86\pm0.15 $      & 3.58$\pm0.75$  & 125 \\
\hline
\end{tabular}
\vspace{-3mm}
\tablecomments{
 $^1$ Column density of the ionized absorber in units of 10$^{22} {\rm atom\, cm^{-2}}$. 
 $^2$ Power-law normalization at 1 keV, in units $10^{-3} $ photons cm$^{-2}$ s$^{-1}$. 
 $^3$  The normalization of the scattered continuum component at 1 keV, in units $10^{-3} $ photons cm$^{-2}$ s$^{-1}$. 
 $^4$ The Fe K emission line normalization is given in units $10^{-5} $ photons cm$^{-2}$ s$^{-1}$.   
In the fit, photon index, the ionization of the absorber and the column density of the scattering gas were consistent with the same values for all observations and were therefore linked in the fit. This yielded:  log $\xi=2.31\pm0.02$, 
$\Gamma=2.33\pm0.05$ and a  column density of the neutral scattering gas  ${\rm N_H} = 6.0\pm 3.1  \times 10^{24} {\rm cm^{-2}}$.  A reprocessed Fe K$\alpha$ emission line complex was allowed in the fit, with column and inclination fixed to those of the scattering gas.  $\chi^2_r=1.06/537$ dof. Errors are calculated at 90\% confidence.
}
\label{tab:params}

\end{table*}%

To remain within the well-calibrated regime for {\it NuSTAR}, this paper restricts
consideration of spectral data to the $\sim$2 - 70 keV band. We rebin the spectral data
first to 1024 channels to reduce the spectral oversampling, then apply
a grouping to achieve a minimum of 20 photons per energy channel in
order to facilitate the use of $\chi^2$ statistics.  Models were fit
to the data using the software package {\sc xspec} ver. 12.9.0 (Arnaud
1996). All models included the Galactic line-of-sight absorption,
N$_{\rm H,Gal} = 1.35 \times 10^{20}{\rm cm}^{-2}$ \citep{dickey90a},
although this had a negligible effect on the fit parameters owing to
the spectral cut-off being relatively high, at 2 keV.  All model
components were adjusted to be at the redshift of the host galaxy,
except for the Galactic absorption. In the following, unless otherwise
stated, fit parameters are quoted in the rest-frame of the source and
errors are at the 90\% confidence level for one interesting parameter
($\Delta \chi^2 = 2.706$).

Initial inspection of the data revealed the source to be flatter in
the low flux states, and steeper in the high states, as observed
previously \citep{miller10a}. Based upon this, and the previous
modeling of the source \citep{miller10a, lobban11a} we adopted a
simple model of a power law continuum plus neutral scattering gas, all covered with an 
ionized absorber plus the Galactic column density of neutral gas. 

While the ionized absorber is known to be complex and
multi-zoned, the lack of soft-band data, and modest spectral resolution of these data prevented us from creating a
sophisticated model of that absorbing gas. It was found that a single ionized gas zone
was adequate to model the low energy behavior of the source. Thus we fit the $2.0 - 70.0$\,keV data using a model composed of a
power-law continuum, modified by passage through a uniform sphere of
ionized gas characterized by an {\sc xstar} model table. For the
ionized absorber table we used version 2.1ln11 of the {\sc xstar} code
\citep{kallman01a,kallman04a}, assuming the abundances of
\citet{grevesse98a}. {\sc xstar} models the absorbing gas as thin
slabs, with parameters of atomic column density and ionization parameter $\xi$,
defined as $$\xi=\frac{L_{\rm ion}}{nR^2}$$ that has units erg\,cm\,s$^{-1}$
and where $L_{\rm ion}$ is the ionizing luminosity between 1 and
1000 Rydbergs, $n$ is the gas density in cm$^{-3}$ and $R$ is the
radial distance (cm) of the absorbing gas from the central continuum
source. The spectral energy distribution was taken to be a simple
power law with $\Gamma=2.5$. Following \citet{lobban11a} the turbulent
velocity was taken as $\sigma=200$ km s$^{-1}$.

The neutral scattering reprocessor gas was modeled using the toroidal
reprocessor model, {\sc MYTorus} \citep{murphy09a, yaqoob10a}.  The
torus has a circular cross section, whose diameter is characterized by
the equatorial column density, ${\rm N_H}$. The torus is illuminated
by a central X-ray continuum. The torus is assumed to have an opening
angle (half angle) of 60$^{\rm o}$, and this corresponds to a global
covering factor of 50\% (and a solid angle subtended by the structure
at the central X-ray source of 2$\pi$ steradians).  The {\sc  MYTorus} model self-consistently calculates 
the Fe K$\alpha$ and Fe K$\beta$ fluorescent emission lines in a separate table to the
opacity profile and the scattered spectrum.  The availability of
these components of reprocessing in separate tables allows the use of
the so-called ``decoupled mode'', whereby one can allow the scattered,
line emission and opacity tables to vary independently, to allow for
time delays between direct continuum, Compton-scattered continuum, and
fluorescent line photons.  All tables include the effects of Compton
scattering.  The element abundances for {\sc MYTorus} are solar
\citep{anders89a} and the photoelectric absorption cross-sections
are those of \citet{verner96a}.
Our model included only the scattered spectrum 
and line emission component.  Our fits assumed an inclination angle of 
0$^{\rm o}$ for {\sc MYTorus}, representative of a face-on 
scattering torus, with the reprocessing matter out of the direct line-of-sight. 

This simple model provided a good fit to all of the flux states exhibited by
NGC~4051 (Fig.\,\ref{fitspectra},  Table\,\ref{tab:params}). The neutral scattering gas yielded
column density ${\rm N_H} = 6.0 \pm 3.1 \times 10^{24} {\rm cm^{-2}}$: with our model construction, this column density 
is that inferred for the scatterer only and not associated with the 
direct transmitted continuum. The emission
line component of {\sc MYTorus} has been converted to Fe K$\alpha$
line fluxes and equivalent widths for ease of interpretation of the
tabulated fits.

The data are consistent with a constant column density for the scattering gas, and 
column variations for the ionized absorber such that in the high state, the column of the ionized gas is
low.  Other variations, such as the ionization state or covering fraction of the absorber 
cannot be ruled out
using these data, so our fits provide only a simple parameterization
of that aspect of the source behavior.

Fitting all the {\it NuSTAR} spectra simultaneously with a single spectral model (ie not allowing any spectral variation between 
epochs, so we could obtain the mean scattering fractions),  
we found the scattered X-ray component (including Fe K line) provides a fraction (of scattered to total flux in each band) 
$<0.14$, 0.17$\pm 0.08$, 0.25$\pm0.08$ and 
0.55 $\pm 0.10$ of the total flux in the 2-4.0, 5.0-7.5, 8-15 and 15-70 keV bands, respectively.  The mean spectral model is shown in 
Fig.\,\ref{modelplot_bands}. 

\subsection{Alternative Models}

It is instructive to explore parameter space for reflection models, as inner disk reflection has been invoked previously to explain the presence of negative lags in 
AGN lag spectra. To this end, we adopted the {\sc xilver} model, that represents reflection from Compton Thick gas with fitable ionization parameter \citep{garcia13a}. 
Initially, we fit the spectra using {\sc xilver} plus a continuum powerlaw, allowing a layer of neutral absorption, fixed to account for the Galactic column as in the previous fits. In the fit, the Fe abundance was fixed at the solar value, and the ionization parameter was allowed to be free. The fit allowed the relative fractions of each of these components to vary between flux states. This fit resulted in $\chi^2_r=1.18$ for 504 degrees of freedom. The photon index was $\Gamma= 2.19\pm 0.03$, with ionization parameter $\xi=1.11\pm0.02$. The reflection fraction was 7\% in the 2 - 4 keV band, 23\% in 5 - 7.5 keV, 31\% in 8 - 15 keV and 52\% in the 15-70 keV band and the uncertainties on all bands were $\sim 20\%$.  
Next we allowed the Fe abundance to be free, and then convolved an unconstrained {\sc kdblur} kernel, to account for relativistic blurring. Neither of these additional  freedoms resulted a significant change in the reflection fractions, the soft-band fraction is mostly constrained by the hardness of the spectrum.

\section{Measurement of the X-ray Time-delay Transfer Function}\label{timedelays}

To measure time delays between bands of differing photon energy, we adopt 
{\refbf an extension of} the
method described by \citet{miller10a, miller10b}. In that method, time series
in two broad bands of photon energy were created, then a maximum likelihood method 
was used to
fit a joint model to the power spectral density (PSD) 
in the two bands and to the cross-spectral density,
fitting to the data autocorrelation in the time domain. Time delays
as a function of temporal frequency were obtained from the phases of
the cross-spectral density. The method 
rigorously accounted for gaps in the timing data and allowed accurate estimates of
statistical uncertainties in PSD, cross-spectral density and time delays,
including estimation of the covariance between the measured time lag values if required.
The method has been independently coded and tested by \citep{zoghbi13a}.

In this paper, given the very broad bandpass afforded by the {\em NuSTAR}
observations, we {\refbf wish to measure} the time delays between four energy
bands: 2--4\,keV, 5--7.5\,keV, 8--15\,keV, 15--70\,keV. 
{\refbf In section\,\ref{discussion}
we consider simple models that are jointly fitted to the time lags between 
pairs of these bands, and for that process to be statistically rigorous we need
to estimate the full covariance matrix, not only between the time lag values in
each lag spectrum, but also between the lag spectra of each pair of energy bands:
because energy bands are in common between the possible combinations of pairings
they are correlated.  To measure the covariance, the \citet{miller10a, miller10b}
analysis was extended to jointly fit to three energy bands, thus obtaining
the lag spectra of three possible pairs of energy bands (i.e. the lags between bands
1 \& 2, between 2 \& 3 and between 1 \& 3, out of 3 bands) and the covariance matrix.
Despite other studies in the literature also carrying joint analysis of multiple
bands \citep[e.g.][]{kara15a}, no previous study has taken into account the covariance
between the measurements: those studies are therefore not able to evaluate the statistical
significance of any lag spectral features that may be found simply from inspection of
either plotted points with error bars or from goodness-of-fit statistics - either of those
approaches should require the full covariance to be included.}

{\refbf 
In this analysis, Fourier bandpowers of frequency width $\Delta\log_{10}\nu=0.2$ were evaluated.}
Each observed photon in the energy bands was given equal weight, such that,
in the absence of significant background flux, the optimum signal-to-noise
ratio was obtained, irrespective of the source spectrum or
the instrument spectral response. Note that, although the best available photon
calibration has been applied, inaccuracies in photon energy calibration have little
effect on the results in these broad energy bands, and the analysis did not
make use of knowledge of the instrument's spectral response.  However, the
source spectrum and the instrumental
spectral response should be considered when interpreting the results (for example,
although the highest energy band extends to 70\,keV, the results are completely
dominated by photons from the lower end of the bandpass around 15\,keV).

{\refbf
Owing to the complexity of joint fitting to multiple energy bands, our analysis can currently
only be carried out on three simultaneous energy bands, which leads to the generation of
three, correlated lag spectra. In our discussion, we concentrate on the three bands
$2-4$\,keV, $8-15$\,keV and $15-70$\,keV, although we also consider the lag spectra that
arise from joint analysis of the three bands $2-4$\,keV, $5-7.5$\,keV and $15-70$\,keV, where
the middle energy bands includes the region of Fe\,K emission and absorption.
}

\begin{figure}
\begin{center}
\hbox{
\psfig{figure=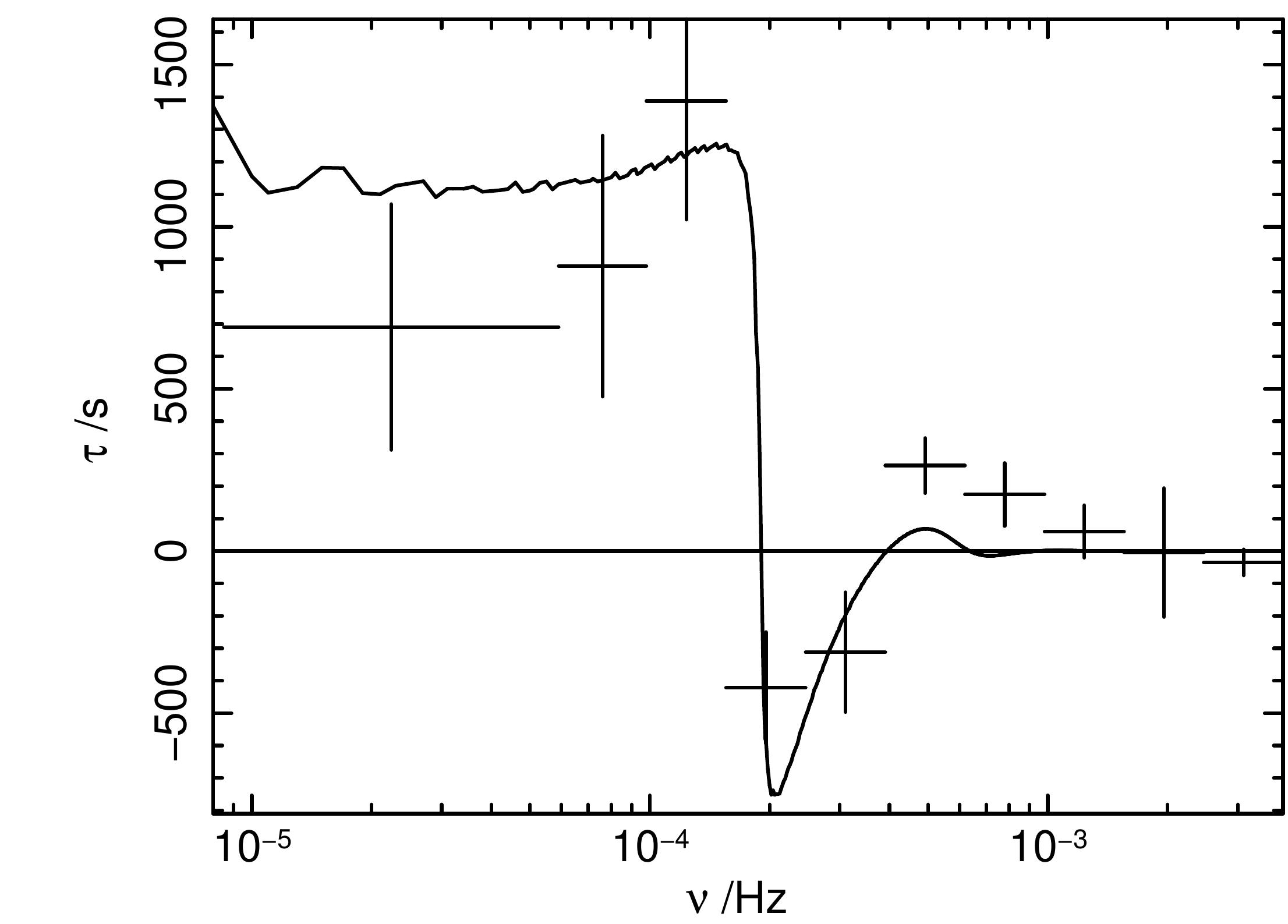, width=1.65in, angle=-90}
}
\hbox{

\psfig{figure=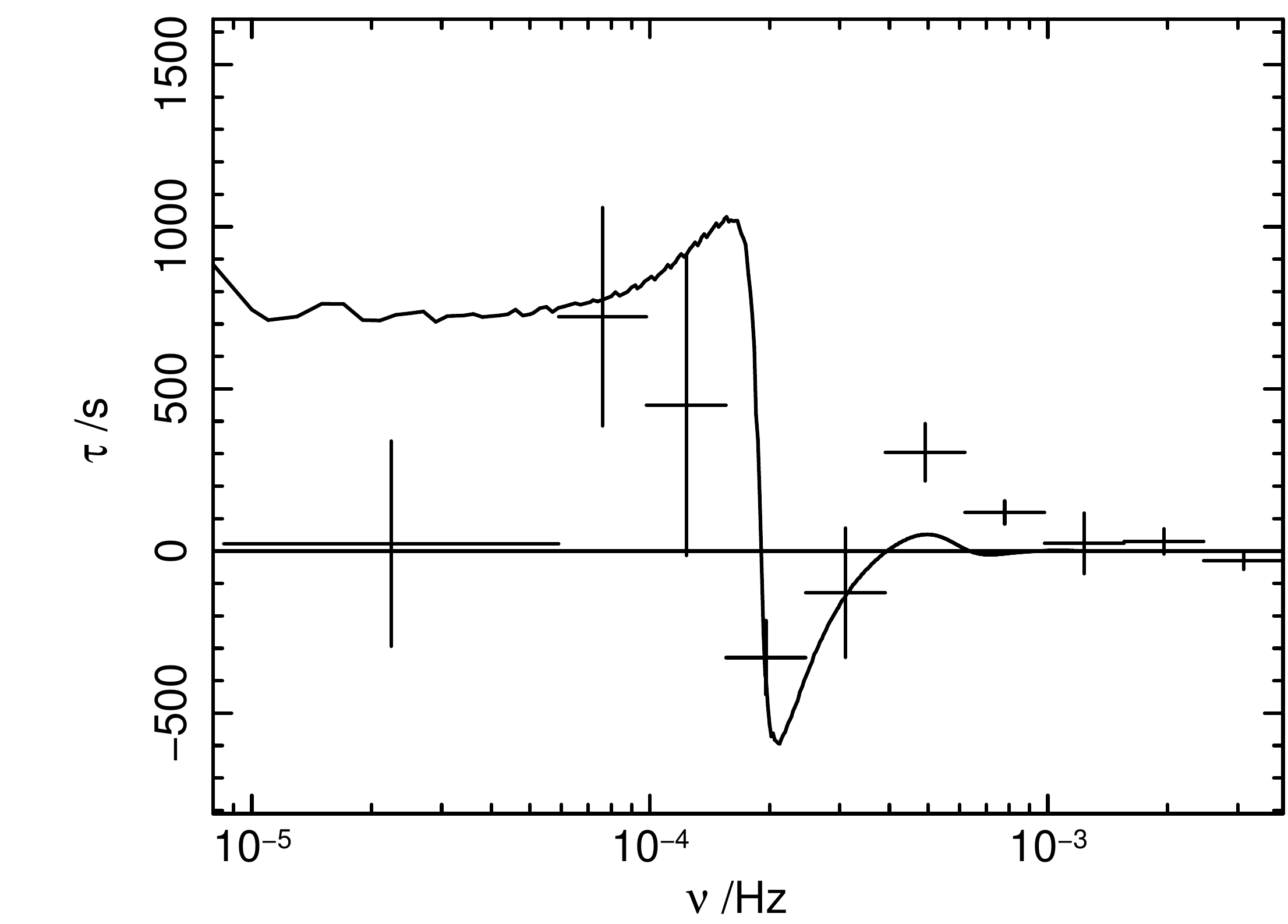, width=1.65in, angle=-90}
}

\hbox{
\psfig{figure=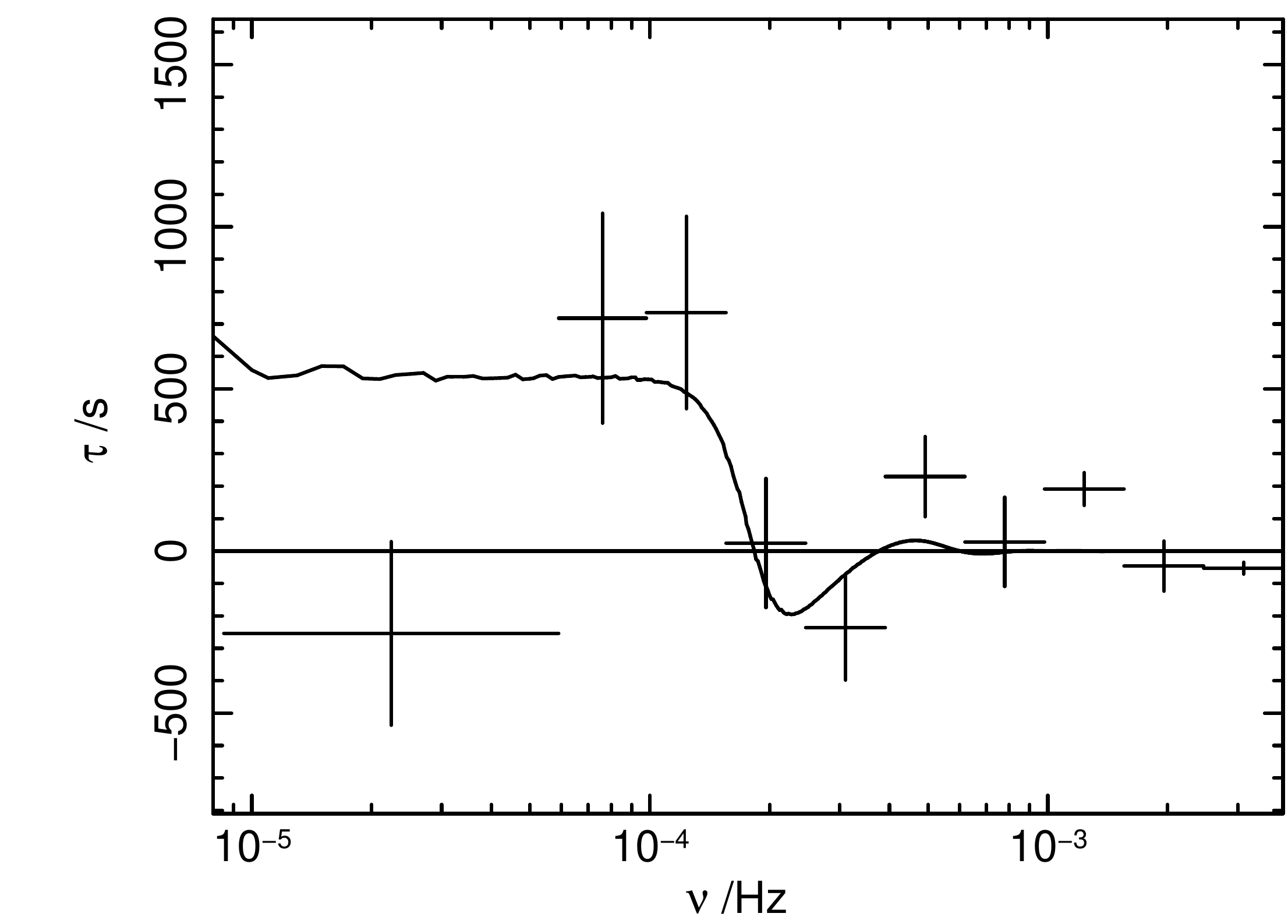, width=1.65in, angle=-90}
}

\hbox{
\psfig{figure=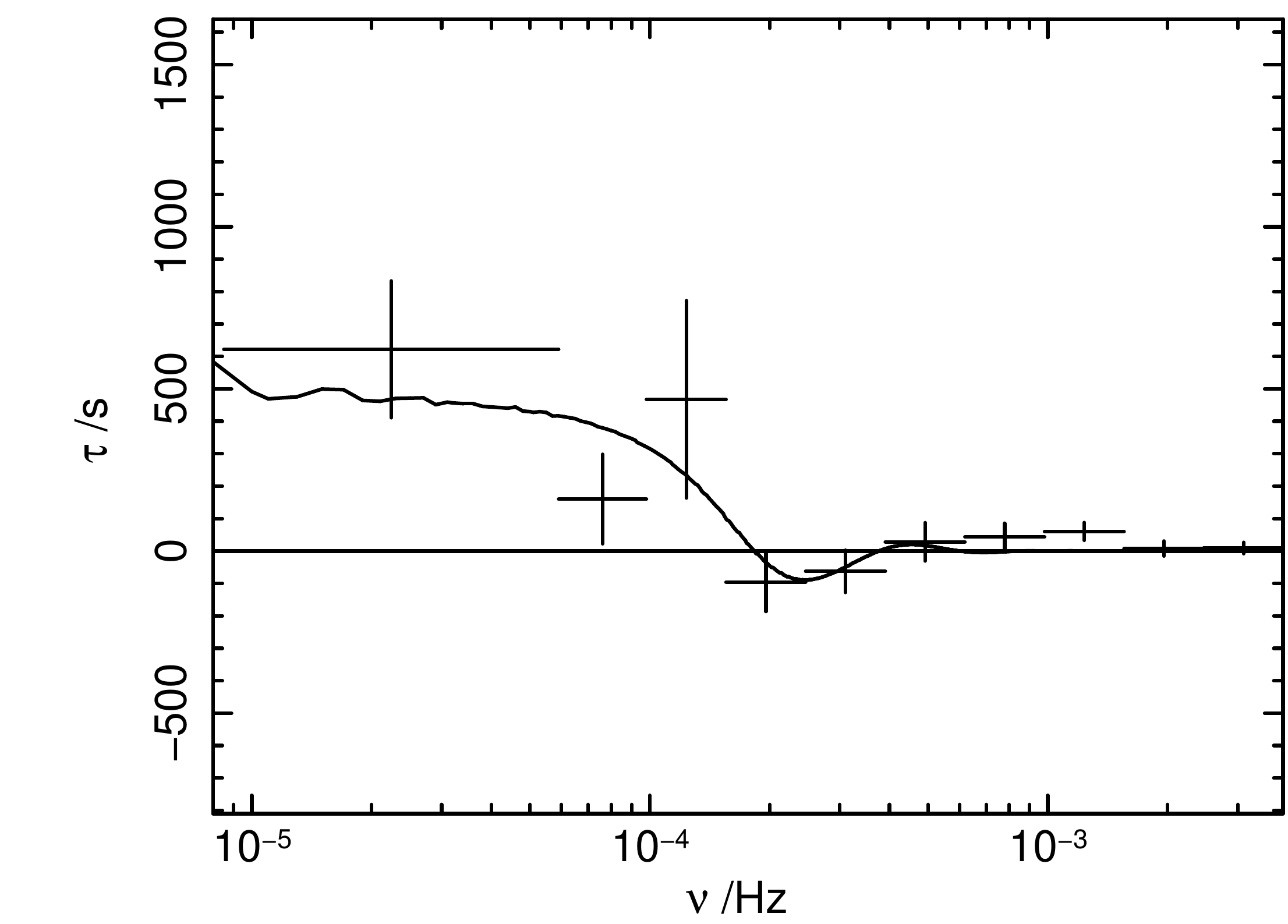, width=1.65in, angle=-90}
}

\hbox{
\psfig{figure=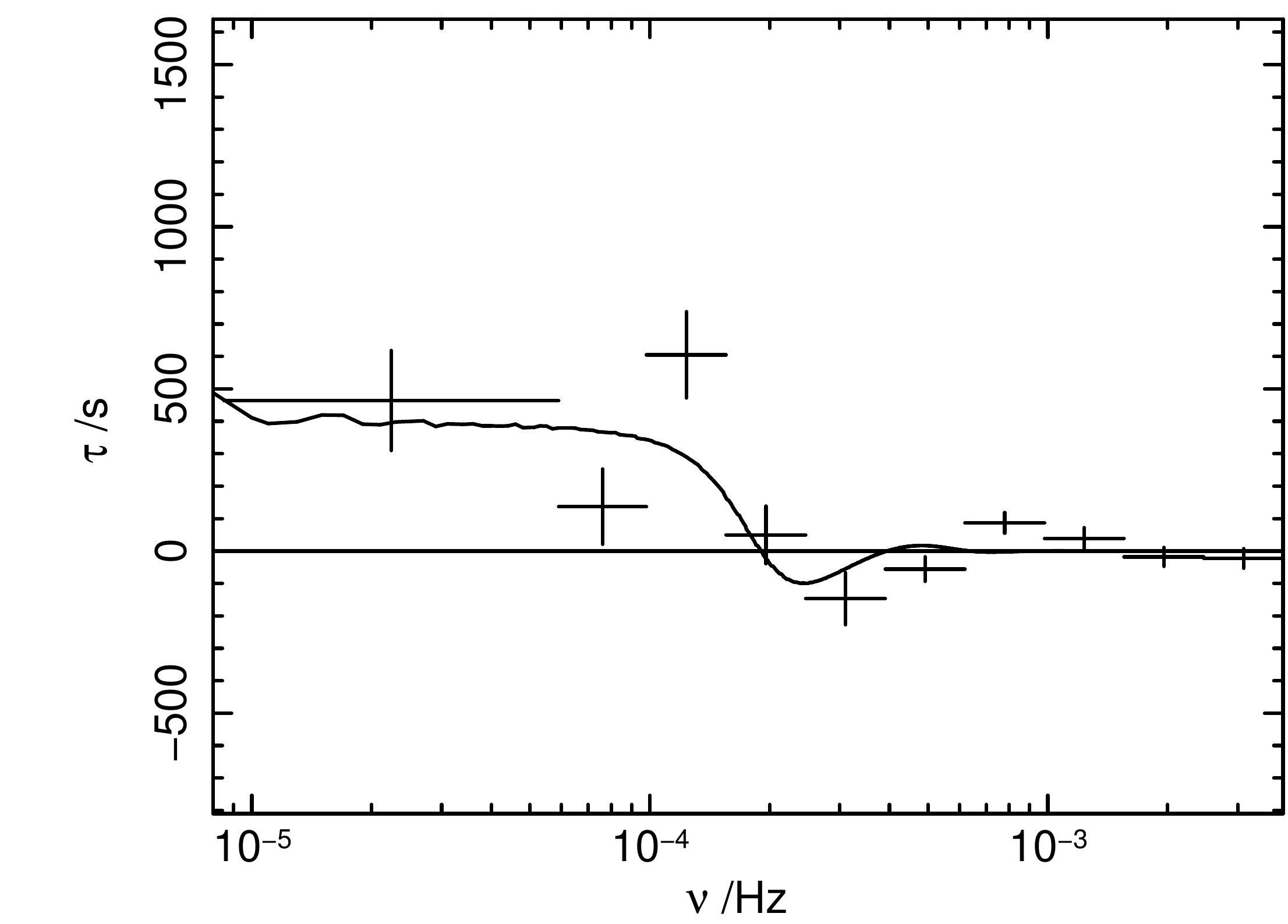, width=1.65in, angle=-90}
}

\end{center}
\caption{Broadband lag spectra of NGC\,4051 from the archived {\it NuSTAR} data, 
shown by the points with error bars,
with the simple top hat response
function described in section\,\ref{transferfunc} (solid line), jointly fitted to all lag spectra.
The panels show delays measured between 
 2-4\,keV \& 15--70\,keV (top),
 5--7.5\,keV \& 15--70\,keV (second),
 8--15\,keV \& 15--70\,keV (third),
 2--4\,keV \& 8--15\,keV (fourth).
 2--4\,keV \& 5--7.5\,keV (bottom).
Note that the plotted points are statistically correlated, both within each lag spectrum and between
lag spectra.
\label{lagspectra}
}
\end{figure}

The `lag spectra' of time delays in all four possible pairs of bands are shown in Fig.\,\ref{lagspectra}. 
The points with error bars show the results from the data analysis, the solid
curve depicts a simple model which will be described in section\,\ref{discussion}.
The horizontal bars on each point show the range of temporal frequency covered,
the vertical bars indicate the statistical uncertainty in the measured values.
In these spectra, we follow the convention that a delay of the higher energy
band with respect to the lower energy band has a positive sign. Thus in all
pairs of energy bands, the harder band consistently lags the softer band at
temporal frequencies around $5 \times 10^{-5}$\,Hz.  The largest time delays have
values around 1000\,s, but some negative delays are also measured with amplitudes
as large as 400\,s. Note that lags that are consistent with zero, or with being slightly
negative, are observed at the lowest frequencies in some combinations of energy
bands, as found also in low states of NGC\,4051 by \citet{alston13a} and as
seen in general in the study of \citet{demarco13a}. Note also the very sharp
negative features visible in the upper two panels of Fig.\,\ref{lagspectra},
with some evidence for such a feature also in the third panel, 
at frequencies around
$2-4 \times 10^{-4}$\,Hz. 

{\refbf
Much attention in the literature has focussed on similar negative lag features, where the
soft-band time variations lead the hard-band time variations in some range of frequency.
Armed with the full covariance matrix, we have evaluated their statistical significance in
these data as follows.  Let us suppose that, where a negative lag value has been measured, 
the true lag value in fact is zero and the negative lag value arises purely from statistical
noise (both photon shot noise and time series sampling variance). 
Then the statistical significance may be evaluated by measuring the change in $\chi^2$
that results when any negative values are forced to a value of zero, instead of being allowed
to be free parameters in the maximisation of the bandpower likelihood.  
The change, $\Delta\chi^2$, may be evaluated using the covariance matrix:
\begin{equation}
\Delta\chi^2 = \vec{\Delta\tau}^T C^{-1} \vec{\Delta\tau} \, ,
\end{equation}
where $\vec{\Delta\tau}$ is a vector of modified lag values, $\vec{\Delta\tau}^T$ is its transpose
and $C^{-1}$ is the inverse covariance matrix.  For positive lags, 
$\Delta\tau = 0$, so they make no contribution to $\Delta\chi^2$. For negative
lags, $\Delta\tau$ is set to the measured maximum-likelihood lag value, so that we measure the effect on $\chi^2$
of forcing these values to be zero. The statistical significance may then be evaluated from a table of
$\chi^2$ with the number of degrees of freedom given by the number of ``zeroed'' lag values. 
We note that this test introduces some element of {\em a posteriori} statistics, as we only test the significance
of values which have been selected by the experimenter.  However, those values have been selected not on the basis of 
their apparent statistical significance, but rather simply on the sign of the lag value, irrespective of whether
the amplitude of the lag value is large or small.
Table\,\ref{tab:chisq} shows the frequency ranges over which negative lags are observed
and their statistical significance, for cross-correlations between all the energy bands that have been considered.
It should be noted that, although the covariance between frequency values has been taken into account,
the values quoted for the various pairings of energy bands are not statistically independent.
Negative lags appear to be significant in all cross-correlations with the hardest band ($15-70$\,keV),
and marginally significant also between the $5-7.5$ and $2-4$\,keV bands.
}
These negative lag features will be discussed further in section\,\ref{discussion}.

\begin{table}
\begin{tabular}{|l|l|c|c|c|c|}
\hline
energy & & 2--4 & 5--7.5 & 8--15 & 15--70 \\
/keV   & &      &        &       & \\
\hline
2--4        & $\Delta\chi^2/\nu$ &          & $10.4/4$    & $3.7/4$     & $45.3/4$  \\
            & $p$                &          & $0.034$     & $0.45$      & $3\times10^{-9}$ \\
5--7.5      & $\Delta\chi^2/\nu$ & 0.25-0.61  &           & n/m         & $29.3/4$  \\
            &    $p$             &          &             &             & $7\times10^{-6}$  \\
8--15       & $\Delta\chi^2/\nu$ & 0.15-0.39 & n/m        &             & $7.5/2$     \\
            & $p$                &          &             &             & $0.024$     \\
15--70      &                    & 0.15-0.39 & 0.15-0.39  & 0.25-0.39   &      \\
\hline
\end{tabular}
\caption{
{\refbf
Table of statistical significance of negative lag values, for combinations
of pairs of energy bands used in the cross-correlation analysis.  An entry of ``n/m''
indicates that lag spectrum was not measured.  The lower left triangle shows the
range of frequency over which a negative lag was observed, in units of mHz. The upper right triangle
gives the $\Delta\chi^2$ value, the number of degrees of freedom, $\nu$, and the $p-$value
statistical significance.  Note that the values shown are not statistically independent.}
}
\label{tab:chisq}
\end{table}

\section{Discussion}\label{discussion}

\subsection{The cross-correlation function}\label{crosscorr}
{\refbf
The time series lags analysis in this and other papers in the literature that
study X-ray time series of AGN is primarily carried out in the Fourier domain. However, much of the
attention in the literature has been focussed on a time-domain interpretation of the lags,
in which the measured time lags are caused by a time delay in the propagation of X-rays 
that are scattered by material some distance from the primary source, either close to the black hole
\citep[e.g.][]{fabian09a} or more distant \citep[e.g.][]{miller10a}. We expect 
such signals to be localised to a finite range of time delays, comparable to the light crossing time
between source and scatterer. It is instructive to inspect the cross-correlation between energy bands (as is commonly
done in optical reverberation studies, e.g. \citealt{peterson04a}) and not
just the lag spectra (which formally are the time lags deduced from the phases of the
Fourier transform of the cross-correlation function).
}

\begin{figure}
\begin{center}
\hbox{
\psfig{figure=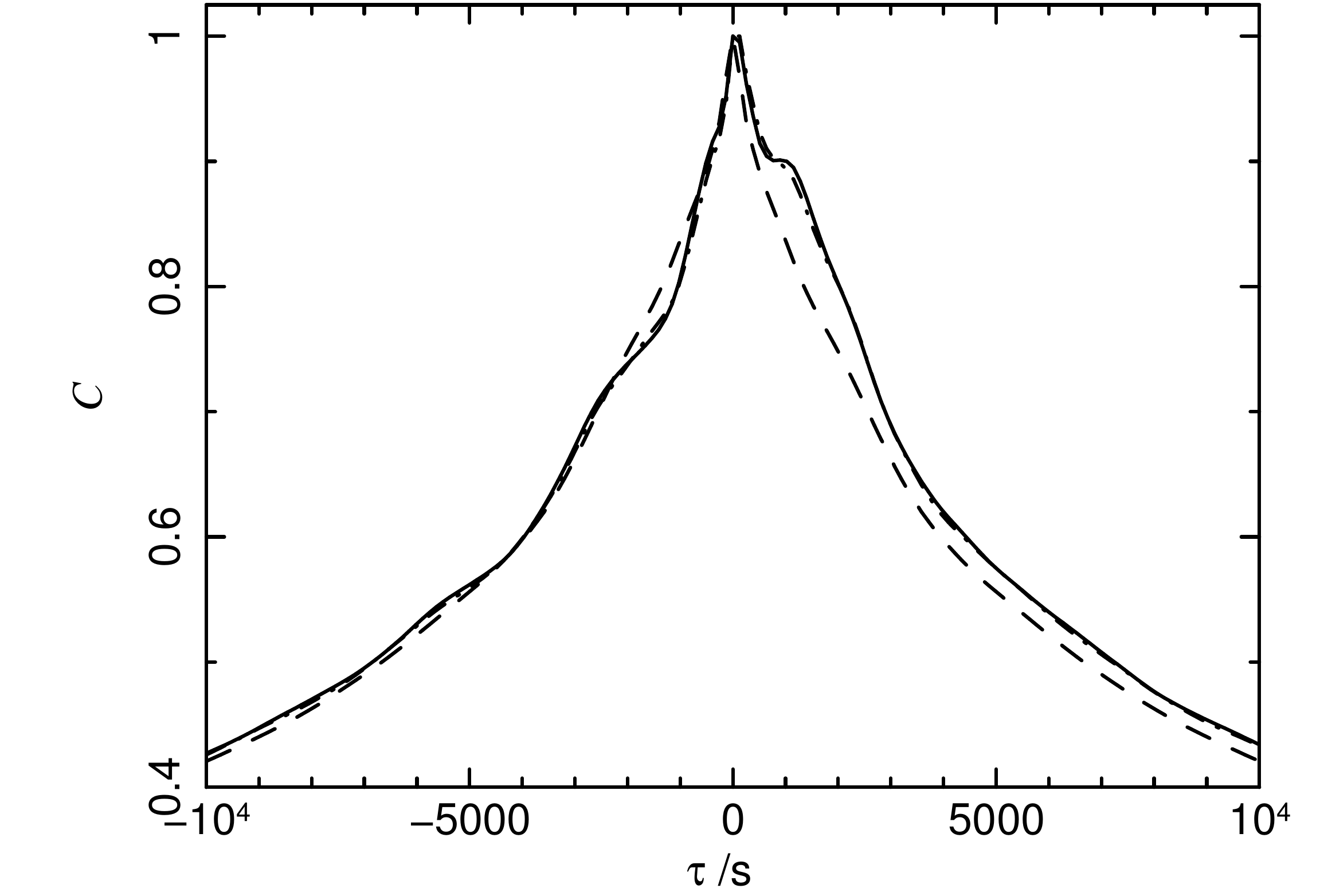,width=2.0in, angle=-90}
}
\end{center}
\caption{
\label{fig:cc}
The cross-correlation function $C$, between the $2-4$\,keV and $15-70$\,keV bands, over a range of time delays
$\tau$ of $\pm10\,000$\,s, 
as deduced in section\,\ref{crosscorr} from the Fourier transform of the maximum-likelihood 
cross spectrum (solid curve).  
The dashed curve shows the symmetric function that arises if all cross-spectrum time lags are
set to zero, as a reference.  The dot-dashed curve (barely distinguishable from the solid curve) shows the 
function that arises if any negative time lags in the cross spectrum are set to zero, with positive time lags
retaining their measured values.
}
\end{figure}

{\refbf
For this inspection, we construct the cross-correlation by Fourier transforming the output cross-spectrum
from our analysis (i.e. utilising both cross-spectrum amplitude and phase information).
As the amplitudes and phases are computed in bandpowers of finite width in frequency, we linearly interpolate
between these bandpowers in order to avoid discontinuities in the cross-spectrum.  The central
$\pm 10\,000$\,s of the resulting cross-correlation function is shown as the solid curve in Fig.\,\ref{fig:cc}.
If the time lags all were zero, the cross-correlation function would be symmetric, as shown by the dashed curve,
which has the same cross-spectrum amplitude as the solid curve.  The most noticeable asymmetric feature is the
shoulder that appears in the range around $1000-3000$\,s, although there is also asymmetric excess at larger
positive time lags.  This shoulder is most naturally interpreted as arising from a delayed excess of emission in
the hard band.  This interpretation is reinforced by the the striking similarity with the flare response functions
measured by \citet{legg12a}: in that study, the temporal shape of flares in emission were obtained by identifying
individual flares in the time series of each band and stacking them. 

Thus the most striking feature of the cross-correlation function may be directly associated with a delayed emission
response in the hard band, as proposed by \citet{legg12a} and earlier work. We may now investigate what 
effect the negative lags in the lag spectrum have on the cross-correlation function.  The dot-dashed curve in
Fig.\,\ref{fig:cc} shows the resulting cross-correlation function when all negative lag values are reset to zero.
This curve is barely distinguishable from the solid curve, with the only visible difference arising as a greater
sharpness of the shoulder feature.  This concurs with the point made by \citet{miller10b} and \citet{miller11a}, 
and discussed further below,
that the negative lags are merely a ringing feature in the Fourier transform of a response function 
that is localised in the time domain.  
}

\subsection{The interpretation of negative lags}\label{neglags}

\subsubsection{The relationship between the time and Fourier domains}
{\refbf
The inspection of the maximum-likelihood cross-correlation function emphasises the point that,
in a time delay analysis such as presented in section\,\ref{timedelays}, it
is essential to jointly consider the full set of energy bands being analysed
and the full set of temporal frequencies.  This is particularly important when,
as seen here, sharp features with negative lag values are seen in the mid-ranges
of the measured frequency range. A sharp feature in Fourier frequency space corresponds to a 
temporally broad, sinusoidal feature in the time domain, and although there may be 
some physical mechanism that could generate such narrow-band features, time lags introduced
by reverberation travel-time delays should instead be limited to a finite range in time delay
and thus produce broadband signals in the Fourier domain. Furthermore, a time domain
feature with a sharp cutoff at some value of maximum time delay naturally produces
oscillations in the Fourier domain.

Thus there are essentially two competing explanations in the literature for such negative lags:
either there is some frequency range in which the soft band flux variations of the source
lag the hard band variations (we follow the standard sign convention that positive lags 
are produced when the harder energy band lags the energy softer band); or, the negative lags are
oscillatory, ringing features caused by taking the Fourier transform of a sharp feature in 
the Fourier domain.  in the remainder of section\,\ref{neglags} 
we discuss the constraints that the analysis may place on these two
interpretations, and we discuss simple models of the second explanation in Section\,\ref{transferfunc}.
}

\subsubsection{Inner disk reverberation coupled with propagating fluctuations in an accretion disk}
{\refbf
An interpretation commonly proposed (e.g. \citealt{zoghbi10a}) is that
there are two separate origins for positive and negative lags.
Low frequency, positive lags are proposed to arise from the inward
propagation of accretion disk fluctuations: changes in harder bands
emitted from close-in are delayed with respect to softer bands from
further out.  The higher-frequency negative lag results from ionized
reflection from an inner accretion disk, whereby the soft band lags
the hard as a result of an anomalously greater contribution of the reflected
spectrum below 1\,keV, compared with a harder reference band
(e.g. above 2\,keV). The combination of these two phenomena is then
suggested to give the oscillatory nature of the lag spectrum (positive
followed by negative lag with increasing frequency).  Such a hybrid model was
proposed for the negative lags observed in 1H\,0707--495 by
\citet{fabian09a} and \citet{zoghbi10a}, and it was proposed that the
strong soft-band reflection contribution arose from an anomalously large excess of Fe\,L
emission in the accretion disk at energies below 1\,keV, caused by
super-solar (by a factor 9) abundance of iron. This model has been discussed
further by \citet{miller10a} and \citet{zoghbi11b}.

Such an explanation does not provide a viable model for the NGC\,4051
data analysed here. The softer reference bands (2--4\,keV, 5--7.5\,keV
or even 8--15\,keV) have a much lower fraction of reflected/scattered
X-rays compared to the hardest (15--70\,keV) band, where most of the
Compton hump is apparent (Fig.\,\ref{modelplot_bands}), and none of
the energy bands observable by {\it NuSTAR} include emission from
Fe\,L ions as had been proposed by \citet{fabian09a} for
1H\,0707--495. There does not seem to be any reverberation mechanism
by which the softer {\em NuSTAR} energy bands could produce soft-band
time variations that lag the hardest energy bands.

This point is reinforced by the observation that the mid-frequency
negative lags detected here are seen in multiple combinations of energy
bands.  In fact, they are seen whenever the hardest energy band is included
in the analysis, but not when only soft bands are included.  This observation
lends weight to the argument that the negative features are not associated with
an excess of delayed, soft band emission.
}

\subsubsection{Ringing features in the Fourier domain}
A significant cause of the lack of consensus to date on the most likely 
physical explanation is that measurement of the 
relative time delay between energy bands by a cross-correlation technique does not
reveal whether the time lag arises in one band or the other, or in some combination
of the two. 
One analysis of particular relevance to this issue was made by \citet{legg12a},
{\refbf already discussed briefly in section\,\ref{crosscorr}.
In that work,} the time series in individual energy bands for three AGN, including 
the {\it Suzaku} observations of NGC\,4051 (section\,\ref{background}), were analysed
assuming a `moving average' model for the response function. In one AGN
in particular, Ark\,564, a very clear signal was obtained, showing
a sharp-edged response function extending a few ks after the primary flares
in emission at energies above 4\,keV, with no evidence for such a response
at lower photon energies. 
So in Ark~564  a reverberation-like signal has been measured directly, 
as a lag in the hard band, not the soft band \citep{legg12a}.  The sharp negative lag in
Ark\,564 arises from a sharp-edged temporal response function in the hard band \citep{legg12a}.
Such an explanation is well-suited to the sharp negative feature seen in the present data
for NGC\,4051\footnote{
We note that the {\em Suzaku} data for NGC\,4051 were of insufficient quality
to reveal any sharp features in the lag spectrum, although the {\it XMM}
data analysed by \citet{alston13a} did detect negative lags.
},
as it arises in multiple combinations of energy bands, not only those
that might contain a particular spectral feature.

\subsection{A simple response function}\label{transferfunc}
To illustrate the ability of the latter model to fit the NGC\,4051 {\em NuSTAR}
results, we have fitted a simple top-hat response function 
jointly to the time lags in sets of three pairs of energy bands. The response function
follows that discussed for 1H\,0707-495 by \citet{miller10b}.  In each band,
the response function comprises a delta function at the time origin, corresponding to 
a contribution from unscattered (direct) emission, plus a top-hat delayed
function with a minimum and maximum time delay.  
{\refbf
At this stage, we do not need to identify a physical origin for this mathematical
response function, it may be viewed as a basic moving-average model
that is simple enough, with only a few free parameters, to fit to the limited 
time lag data that is available. 
However, the top-hat response function does have a simple interpretation in the context of reverberation models:
a delayed top-hat response corresponds to the reverberation expected from a thin spherical shell of scattering
material.  We discuss this interpretation below in section\,\ref{lightecho}, but here we consider this
simple response function merely as a convenient simple model that may reproduce the observed
lag spectra and allow us to extract some basic parameters that describe the observed frequency-dependent lags
without adopting any particular physical interpretation.
Oscillatory lags in the lag spectra arise as ringing in the Fourier
transform, but only when the minimum time delay is greater than zero, do the oscillations go negative.
}

For this illustration, the simplest response function that was tested comprised 
{\refbf a 'direct' component, i.e. delta function at the origin} 
plus a top-hat delayed response defined to be non-zero over a finite range of time
delay $\tau_{\mathrm min} < \tau < \tau_{\mathrm max}$. The time delays were 
free parameters with
$\tau_{\mathrm min},\tau_{\mathrm max} \ge 0$ and had the same 
value in all energy bands. 
The ratio of the flux in the direct and
{\refbf delayed} components was a free parameter in each energy band.
Thus in fitting to any combination of three energy bands there were
a total of 5 free parameters. 
The lag spectrum values were predicted from this response function and a joint best fit to the 
set of measured lag spectra was obtained by mimimising the 
total $\chi^2$ between the predicted values and the measured lag spectrum points,
{\refbf including in the $\chi^2$ calculation the covariance between the measured 
time delay values}. The response function-fitting was repeated for differing combinations of
three of the four energy bands used in our analysis and consistent results were
obtained.  
{\refbf Fitting was carried out using a Markov Chain Monte Carlo (MCMC) sampling procedure, where
convergence of the MCMC chains was established by inspection of the convergence of the distributions
resulting from 24 independent chains with differing starting positions.
}

\begin{figure}
\begin{center}
\hbox{
\psfig{figure=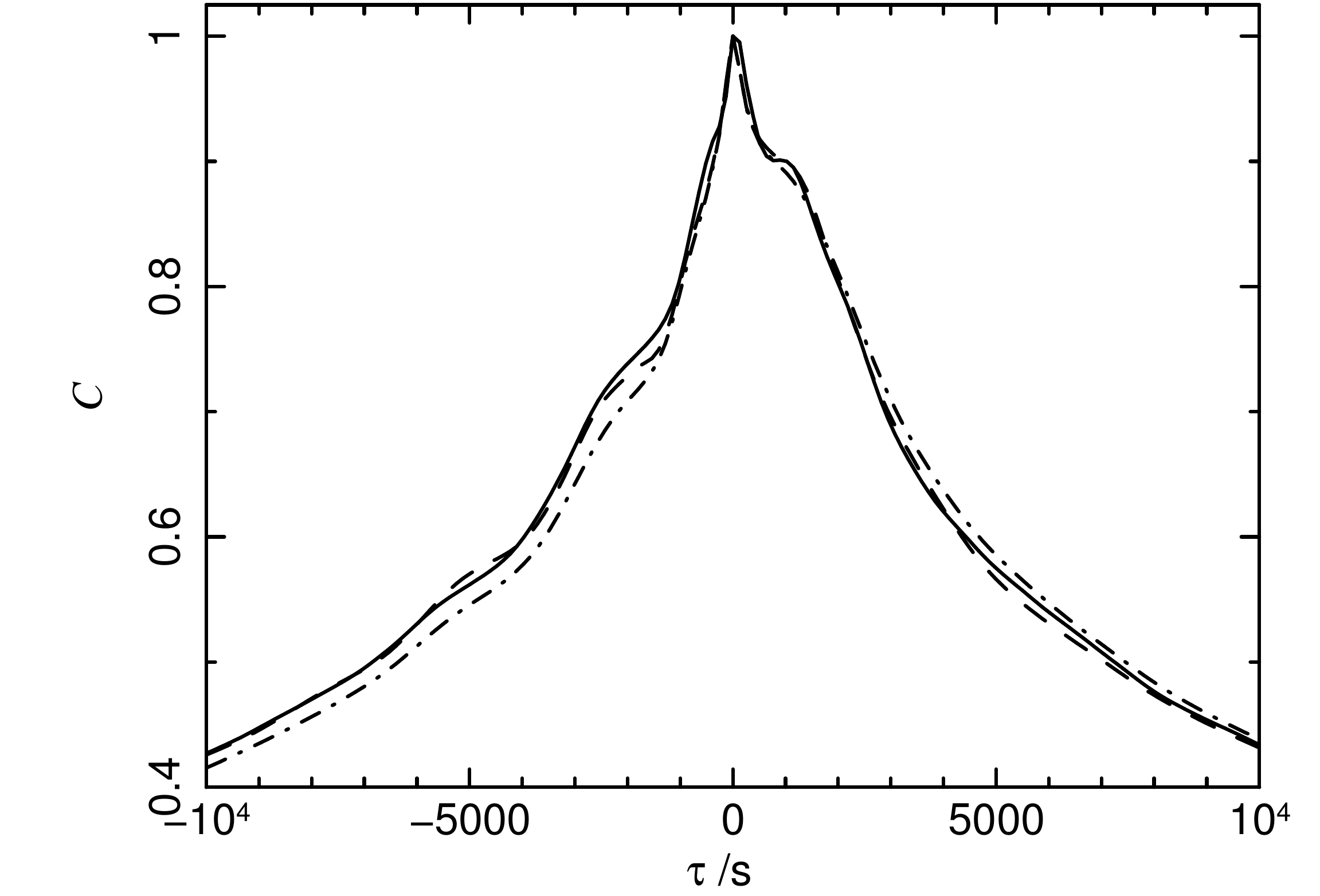,width=2.0in, angle=-90}
}
\end{center}
\caption{
\label{cc_comparison}
The cross-correlation function $C$
as deduced in section\,\ref{transferfunc} using the time delays of the
simple top-hat response function (dot-dashed curve) compared
with the full cross-correlation function of Fig.\,\ref{fig:cc} (solid curve).
Also shown is the more complex response function discussed in
section\,\ref{transferfunc}.
}
\end{figure}

{\refbf 
Fig.\,\ref{lagspectra} shows the lags predicted from this best-fit, simple top hat response
function superimposed on the measured lag spectrum values.
This response function yields a qualitatively reasonable fit to the lag spectra, and in particular
is able to reproduce well the large positive lags at low frequency and the negative lags at mid
frequencies. It is also able to reproduce the dominant `shoulder' feature in the cross-correlation
function, shown in Fig.\,\ref{cc_comparison} as the dot-dashed curve, compared with the measured
cross-correlation function, shown as the solid curve.
However, the overall goodness-of-fit to the lag spectra is poor, with $\chi^2=71, 65$ for the two sets of lag
spectrum analysis (one analysis using the $2-4, 5-7.5, 15-70$\,keV bands, the other the $2-4, 8-15, 15-70$\,keV
bands, respectively, with 25 degrees of freedom in each).  
Inspection of the lag spectra and the largest contributions to $\chi^2$
indicates that this poor goodness of fit largely arises at higher frequencies, and in particular this
simple response function has difficulty in reproducing the positive lags seen in some of the lag spectra
at frequencies $\nu \simeq 5\times 10^{-4}$\,Hz.
The absence of high frequency lags has little effect on this time domain function, however, as these
primarily determine only the detailed shape of the cross-correlation features.

To better fit the complexity of the lag spectra, a more complex response function was tested, comprising two such 
top-hat components superimposed. Thus there were four parameters to describe the
$\tau_{\mathrm min}, \tau_{\mathrm max}$ values of these two components, and the ratios of direct
to delayed flux in each component were free parameters in each band, making a total of 10 free
parameters. This more complex response function does fit better, and matches better the
cross-correlation function, shown as the dashed curve in Fig.\,\ref{cc_comparison}. 
However, the improvement in
$\chi^2$ is only $\Delta\chi^2 \simeq 10$ for each lag spectral analysis, 
which given that 5 additional parameters have been introduced, does not make this more 
complex function a compelling choice. The complex behavior exhibited in these data 
may require the addition of a different component than a second top hat function.  However, we defer more complex modeling to future work.

Although higher frequency, positive lags need to be included to obtain a good fit to the
joint set of lag spectra, it seems clear that this class of response functions
is able to explain the gross time delay features that arise at low and mid frequencies, and in particular
the transition to negative lags and mid frequencies.
}
We do not expect that such a simple response function should be correct in detail - 
the physics of the emission and reprocessing region and its geometry must surely be more 
complex than can be described by just five numbers, and the function discussed in this
section is indicative only - but it does show how the most obvious features in the
lag spectra may have a common origin in all the energy bands over the full energy range
of the {\em NuSTAR} data - and that, as seen in Ark\,564, sharp negative lags 
may arise without any delayed response in the softest energy band. 

\subsection{A simple light echo model for NGC\,4051}\label{lightecho}
The above simple top-hat model may be viewed as a conveniently simple
mathematical model of the response function, without any particular physical
interpretation being implied.  However, the top-hat model does have a simple
interpretation in the context of light echo reverberation models, as we now discuss.

A top-hat temporal response function arises in the case of scattered emission from a 
uniform, isotropic, thin shell surrounding a central source, where the maximum
time delay in the response function is given by the light travel time across the
diameter of the shell. A minimum time delay that is greater than zero may arise
if scattered light close to the line of sight is absent, such as might be caused either by holes in 
the scattering distribution, or by the scattering being caused by illuminated surfaces that, 
along the line of sight, face away  from the observer.  Comparison of our lag spectra with those obtained for NGC 4051 from 
{\it XMM} data \citep{emmanoulopoulos14a, demarco13a} show a transition  from positive to negative lag values at higher frequencies ($ 3 - 4 \times 10^{-4}$ Hz) than detected here. As the transition frequency relates to the diameter of the reverberating shell, those data imply reverberation from a  shell  a factor $\sim 2$ smaller than that observed during the {\it NuSTAR} epoch.  

{\it XMM} data also provide evidence that a zone of reprocessing gas which is not in equilibrium can result in  a delay between the more absorbed photons and the continuum photons, producing a negative lag in the source lag spectrum \citep{silva16a}: those authors found soft lags of $\sim 100$s for temporal variations on timescales of hours. The zone of gas producing that lag is implied to exist more than an order of magnitude further from the active nucleus than the reverberating shell discussed here. Owing to the lack of soft-band data using {\it NuSTAR}, we cannot investigate the possibility of warm absorber induced components of lag in the data presented here. 

\begin{table}
\begin{center}
\begin{tabular}{lllllll}
\hline
band & t$_{\rm min}$/s & & t$_{\rm max}$/s & & $R$ & \\
\hline
2-4\,keV & $1845^{+520}_{-560}$ & & $3715^{+780}_{-720}$ & & $0.015^{+.027}_{-.005}$ \\
5-7.5\,keV &                     & &                     & & $0.28^{+.1}_{-.1}$ \\
15-70\,keV &                     & &                     & & $0.61^{+.12}_{-.36}$ \\
\hline
\end{tabular}
\end{center}
\caption{Parameters of the best-fit top-hat response function for the
joint fit to the three energy bands $2-4$, $5-7.5$ and $15-70$\,keV.
The minimum time delay (column 2) and
maximum time delay (column 3) were constrained to be the same in all three
bands. The fraction of scattered light, $R$ (column 4) is given in each energy
band. Uncertainties denote 68\,percent confidence intervals.}
\label{tab:bestfitmodel}
\end{table}

The parameters of the best-fit simple response function from Section\,\ref{transferfunc}
are shown in Table\,\ref{tab:bestfitmodel}. 
It may be seen that the fraction of scattered light, $R$, increases with increasing energy, as expected
in the scattering of X-rays from absorbing material, either an accretion disk
or more general circumnuclear material \citep[e.g.][]{miller13a}.  These scattered fractions are consistent with those found from the spectral model (Section 4). 

The scattered fractions observed in these data (Table\,\ref{tab:bestfitmodel}) are 
consistent with those of \citet{miller13a}. In those models, continuum emission
from the primary source is suppressed by partial covering of material with 
geometrical covering fraction around 50\,percent that has a 
high optical depth to Compton scattering, that emission is replaced by time-delayed scattered light from the 
illuminated surfaces of the circumnuclear material. 
Clearly, the current observations and analysis are incapable of definitively testing such
a simple model, but future modelling may be able to provide testable predictions
of the light echo scenario. 
We make one final note here, that the light echos expected from
the \citet{miller13a} models may also predict an excess of delayed emission in the `red wing'
continuum region at energies below Fe\,K, as seen in some other AGN \citep[e.g.][]{kara15a}.

\section{Conclusions}

Timing analysis of five archived {\it NuSTAR} observations of NGC~4051 show lags between flux variations in different energy bands.  
The harder band flux variations consistently lag the softer band, with lags of at least 1000\,s, at 
temporal frequencies $\nu \sim 5 \times 10^{-5}$ Hz. The data also show 
statistically significant negative lags, i.e. soft-photon lags  up to amplitudes of 400s at temporal frequencies $\nu \sim 2 \times\, 10^{-4}$ Hz, particularly when the highest photon energies are used in the
cross-correlation.

The presence of a negative, soft lag in the cross correlation between bands where the softer energy band should contain little or no scattered light indicates that negative (soft band lags the harder band) lags do not arise from soft band scattered or reflected light during the {\it NuSTAR} observations. This argues against an origin of the negative lag as a reverberation signal arising from an excess of soft band reflection 
from the inner accretion disk, as has been suggested for similar timing behaviour observed in other AGN. 

The hard band time delays at low frequencies and these soft band delays at mid frequencies may be reproduced
with very simple top hat response functions, that may be used to quantify the delay timescales in the system.
It has been shown that negative lags may arise from ringing effects in the Fourier transform of a
delay in the hardest bands. Inspection of the time domain cross-correlation function reveals a distinct
shoulder on timescales of 1--3\,ksec which are consistent with arising as a delay in the hardest energy
bands.  It has been demonstrated that the negative lags in the Fourier domain have little effect on the
cross-correlation function, other than to increase the sharpness of the shoulder feature.  There 
is good agreement with the time domain delayed features seen in the direct stacking of X-ray flares by \citet{legg12a}.

These observations may all be produced by reverberation, in which X-rays are scattered by material 
with light travel times a few thousand ksec from the primary X-ray source, corresponding to a few hundred
gravitational radii.  Using the parameterisation of the simple top hat
response function, scattering fractions derived from the data indicate the reprocessor 
to have a global covering fraction $\sim 50\%$ around the primary source. 
The scattered fraction of light increases with increasing photon energy, as expected for scattering by 
photoelectrically absorbing material. 

\section{Acknowledgements}
\label{lastpage}

T.J.Turner acknowledges financial support from NASA grant NNX11AJ57G. 
J.N. Reeves acknowledges financial support from STFC and from NASA grant number NNX15AF12G. VB acknowledges the support from the grant ASI-INAF {\it Nustar} I/037/12/0. This research has made use of data obtained with the {\it NuSTAR} mission, a project led by the California Institute of Technology (Caltech) and managed by the Jet Propulsion Laboratory (JPL).

\bibliographystyle{mnras}      
\bibliography{xray_31jan2017}   

\end{document}